\documentclass[letterpaper]{article} 
\usepackage{aaai2027}  
\usepackage[hyphens]{url}  
\usepackage{graphicx} 
\usepackage{natbib}  
\usepackage{caption} 
\usepackage{algorithm}
\usepackage{algorithmic}

\usepackage{newfloat}
\usepackage{listings}
\DeclareCaptionStyle{ruled}{labelfont=normalfont,labelsep=colon,strut=off} 
\floatstyle{ruled}
\newfloat{listing}{tb}{lst}{}
\floatname{listing}{Listing}

\usepackage{booktabs}
\usepackage{multirow}
\usepackage{makecell}
\title{Collusion-Resistant Image-Agnostic Watermarking for Multi-Screen Shooting}

\author{
	Mingyue Chen\textsuperscript{\rm 1},
	Xin Liao\textsuperscript{\rm 1}\thanks{Corresponding author},
	Yufeng Wu\textsuperscript{\rm 1},
	Han Fang\textsuperscript{\rm 2},
	Xiaoshuai Wu\textsuperscript{\rm 1}
}

\affiliations{
	\textsuperscript{\rm 1}College of Cyber Science and Technology,
	Hunan University, Changsha, China\\
	\textsuperscript{\rm 2}University of Science and Technology of China,
	Hefei, China\\
}

\begin{document}

\maketitle

\begin{abstract}
Screen-shooting poses a significant threat to confidential information protection. While existing screen-shooting watermarking methods enable copyright verification, the copyrighted images carrying the same copyright watermark across different screens often exhibit highly similar and estimable watermark patterns. These shared patterns can be exploited for watermark removal and forgery, a threat we term the multi-screen collusion attack.
To mitigate this threat, we propose \textbf{CoMSMark}, a \textbf{co}llusion-resistant image-agnostic watermarking framework for \textbf{m}ulti-\textbf{s}creen shooting, which reduces shared residual components across screens to resist multi-screen collusion attacks.
Specifically, we incorporate screen ID through a style modulation mechanism, enabling the encoder to generate screen-specific watermark residuals for reliable source attribution.
We further introduce a collusion suppression loss that reduces shared residual components and encourages high-entropy predictions for forged samples, improving resistance to collusion attacks.
Finally, to enable efficient large-scale distribution, CoMSMark employs an image-agnostic encoding paradigm that generates watermark residuals independently of image content.
Extensive experiments demonstrate that CoMSMark effectively resists both collusion-based watermark removal and forgery. It maintains an average watermark accuracy above 90\% under removal attacks while keeping forged-watermark accuracy near 50\%.
Moreover, CoMSMark achieves competitive robustness under diverse screen-shooting conditions, including varying capture distances and angles.
\end{abstract}


\section{Introduction}

The widespread use of display devices and high-definition camera‌s has made screen-shooting an increasingly convenient means of unauthorized content acquisition, posing serious risks of copyright infringement and information leakage. Digital watermarking provides an effective solution by embedding imperceptible information into visual content, enabling copyright verification and source tracing~\cite{zong2014robust,chu2003dct, joseph2020image,sun2020robust,jia2021mbrs,sander2025watermark}. To improve robustness against screen-shooting, numerous robust watermarking schemes~\cite{tancik2020stegastamp,fang2022pimog,wu2026sim} have been proposed to withstand complex distortions, such as perspective distortion, illumination variation, and moiré patterns.

\begin{figure}[!t]
	\centering
	\includegraphics[width=\linewidth]{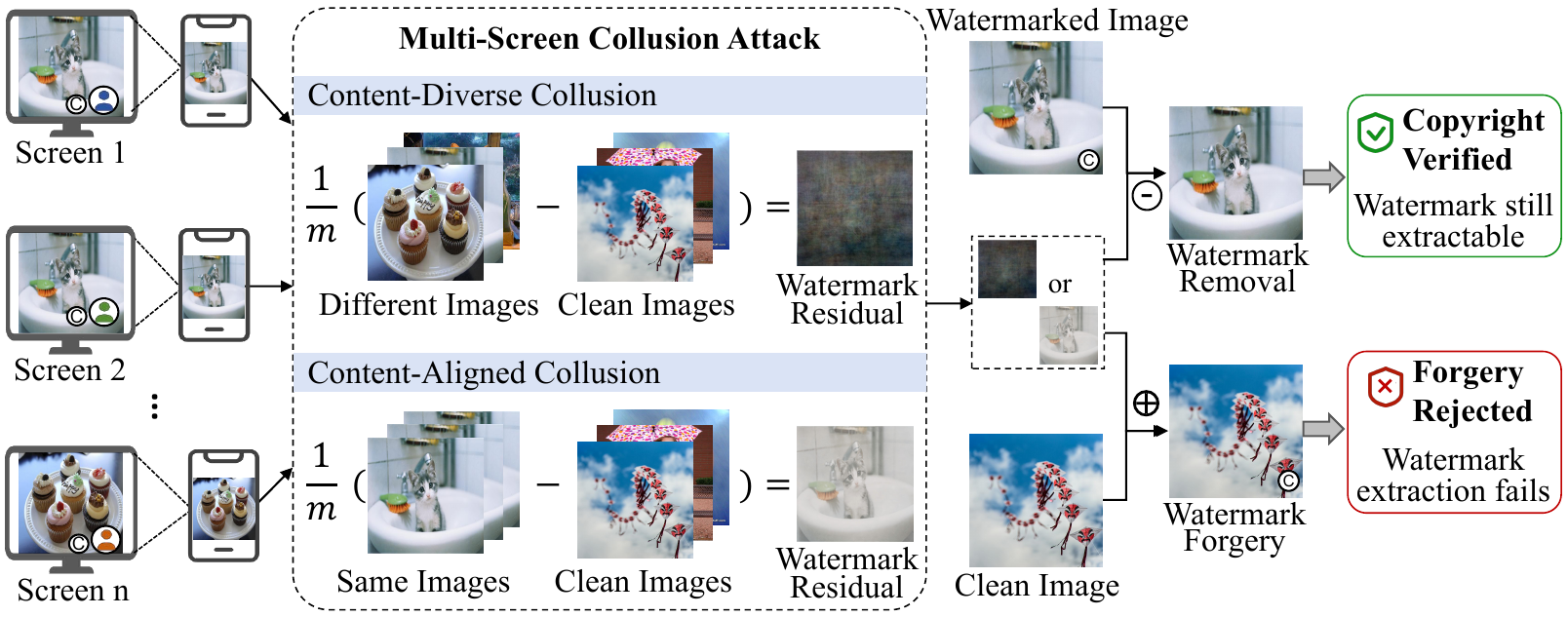}
	\caption{Illustration of multi-screen collusion attacks. Captured images carrying the same watermark can be aggregated under two scenarios: content-diverse collusion, where different images are captured across screens, and content-aligned collusion, where the same image is captured from different screens. The estimated shared watermark residual is then exploited for watermark removal or forgery.}
	\label{fig:first}
\end{figure}

However, existing screen-shooting watermarking methods lack dedicated designs for multi-screen scenarios. When the same watermark pattern is distributed across screens, its shared components pose a risk of multi-screen collusion attacks. 
As illustrated in Fig.~\ref{fig:first}, adversaries can collect captured copies carrying the same copyright watermark from multiple screens and estimate the shared watermark pattern by subtracting independently collected clean images and averaging the resulting residuals~\cite{yang2024steganalysis, fei2026efficient}. 
The estimated pattern can then be used for watermark removal or forgery. Such attacks arise in two practical scenarios. In content-diverse collusion, the captured copies contain different image contents, where content variations can be reduced through aggregation. In content-aligned collusion, the copies share the same image content, providing a more consistent basis for residual estimation under aligned image structures. To the best of our knowledge, this multi-screen collusion attacks has not been explicitly investigated in existing screen-shooting watermarking research.

The main reason existing methods~\cite{tancik2020stegastamp,jia2020rihoop,fang2022pimog,chen2025flexible} remain vulnerable to such attacks is that their watermark residuals exhibit statistical consistency across copies distributed to different screens. A captured watermarked image can be viewed as a combination of natural image content, an embedded watermark residual, and screen-shooting distortions. When captured copies from multiple screens are averaged, image-specific content and capture distortions are attenuated because they vary across samples, whereas the shared watermark component is preserved. Moreover, subtracting independently collected clean images from the aggregated watermarked copies further attenuates the natural-image components, making the shared watermark pattern more distinguishable. Consequently, even without access to corresponding original images, an adversary can estimate a usable watermark residual from sufficiently many captured copies, posing a new security challenge to existing screen-shooting watermarking methods.


Therefore, a multi-screen watermarking framework should preserve a common copyright message while avoiding statistically consistent residual patterns across screens that can be exploited by multi-screen collusion attacks. It should also support distinguishable screen IDs for source attribution without substantially increasing the watermark payload, while ensuring reliable separation of copyright and identity information during extraction. In addition, the encoding process should remain efficient and scalable for large-scale content distribution, rather than repeatedly performing image-aware encoding for each input image.

To address these challenges, we propose \textbf{CoMSMark}, a \textbf{co}llusion-resistant image-agnostic watermarking framework for \textbf{m}ulti-\textbf{s}creen shooting. During encoding, screen ID is introduced as a conditioning variable to generate screen-specific watermark residuals, enabling source attribution while preserving reliable watermark recovery. Moreover, we introduce a collusion suppression loss that suppresses shared residual components across screens and encourages high-entropy watermark predictions for forged samples, thereby improving resistance to watermark removal and forgery. Finally, the residuals are generated independently of image content and can be directly applied to different images, enabling efficient large-scale multi-screen distribution.

The contributions of this paper are summarized as follows:

\begin{itemize}
    \item We introduce a simple yet serious multi-screen collusion attack on screen-shooting watermarking and systematically analyze its mechanism and corresponding defense.
    \item We propose CoMSMark, a collusion-resistant image-agnostic watermarking framework that jointly supports copyright authentication, leakage source attribution, and collusion resistance, while enabling efficient encoding for large-scale multi-screen distribution.
    \item We design a screen-conditioned encoder with style modulation and a collusion suppression loss to generate screen-specific residuals and enhance collusion resistance.
    \item Extensive experiments demonstrate that CoMSMark effectively resists collusive removal and forgery while maintaining strong watermark robustness.
\end{itemize}

\begin{figure*}
	\centering
	\includegraphics[width=\linewidth]{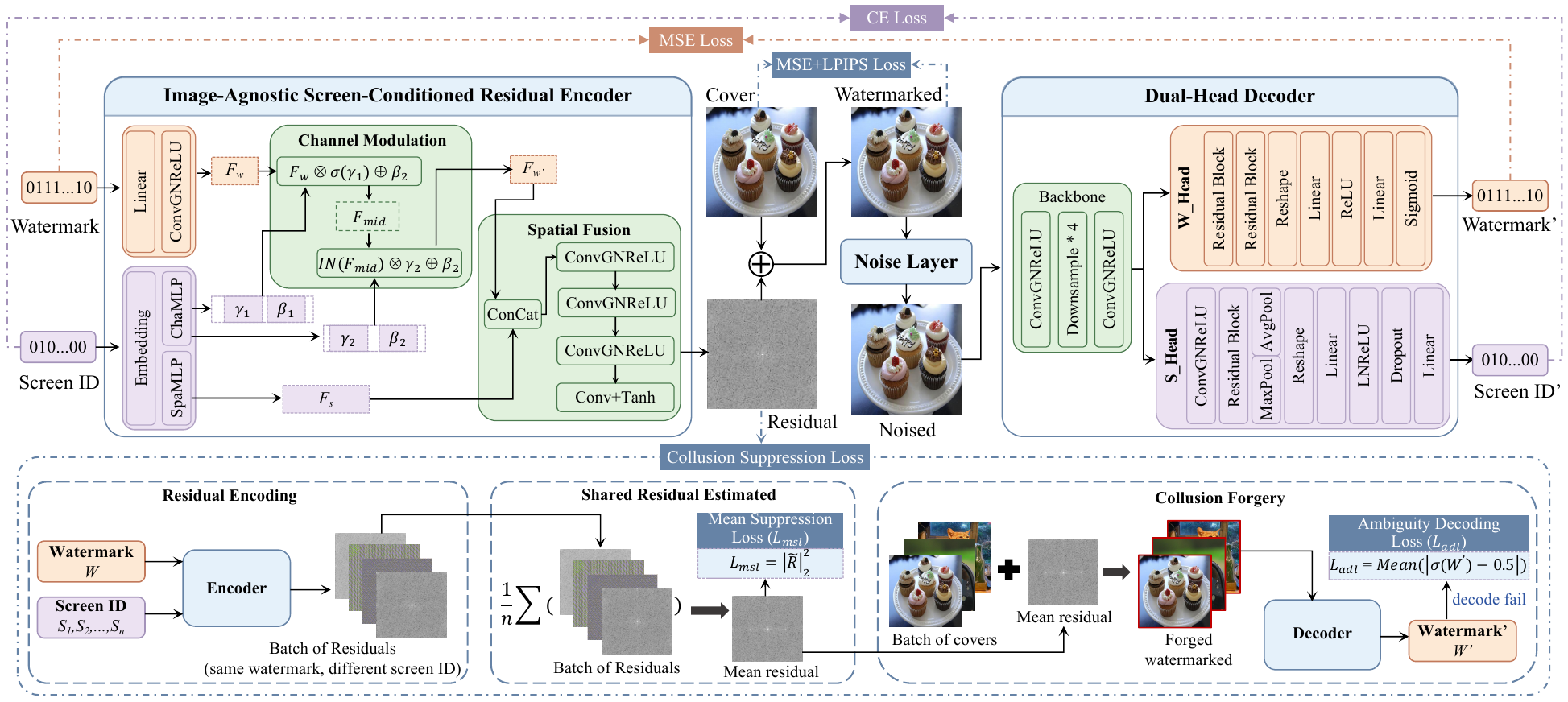}
	\caption{Overview of the proposed CoMSMark framework. The encoder generates screen-specific residuals from the watermark and screen ID, which are added to cover images and processed by a screen-shooting noise layer. A dual-head decoder jointly recovers the watermark and screen ID. The collusion suppression loss reduces shared residual patterns and constructs forged samples during training to enhance resistance to multi-screen collusion attacks.}
	\label{fig:framework}
\end{figure*}

\section{Related Work}
\subsection{Screen-shooting Watermarking}
Screen-shooting watermarking aims to preserve watermark recoverability after visual content is displayed on a screen and recaptured by a camera. Compared with traditional techniques~\cite{fang2018screen,wang2024robust}, deep learning-based frameworks have achieved improved robustness and imperceptibility by modeling the screen-shooting channel with differentiable noise layers~\cite{jia2022learning, li2024screen}. StegaStamp~\cite{tancik2020stegastamp} simulated physical capture through a sequence of differentiable distortions, while PIMoG~\cite{fang2022pimog} modeled perspective transformation, illumination variation, moiré patterns, and Gaussian noise to improve screen-shooting robustness. Subsequent works further explored data-driven distortion modeling and more realistic noise approximation, such as LFM~\cite{wengrowski2019light}, STSR~\cite{gao2025screen}, and S2R~\cite{wu2026sim}. Recent studies also considered partial screen capture: FPSMark~\cite{chen2025flexible} distributed watermark information across multiple regions, and RoPaSS~\cite{ma2025ropass} introduced symmetric watermark patterns for resynchronization under incomplete capture. However, existing methods mainly address screen-shooting robustness but overlook the security risks of multi-screen collusion attacks.

\subsection{Collusion Attacks of Watermarking}
Existing studies on watermark collusion have mainly focused on statistical analysis of multiple watermarked media samples and aggregation of multiple watermarked models.

In digital image watermarking, \cite{yang2024steganalysis} proposed a steganalysis attack that exploits multiple watermarked images to estimate persistent watermark patterns without requiring the corresponding originals. The extracted patterns can degrade watermark verification and enable false watermark claims, while multi-key assignment provides only limited mitigation.

Collusion has also been studied in generative model watermarking, where malicious users combine multiple watermarked model copies to suppress embedded identifiers. \cite{fei2026efficient} introduced user-specific parameter transformations to improve fingerprint robustness against model aggregation, while SWM~\cite{dai2026secure} generated functionally consistent but parametrically distinguishable model variants to reduce the utility of colluded models.

However, multi-screen shooting introduces a new threat scenario where adversaries exploit cross-screen redundancy among captured copies to estimate shared watermark residuals. This scenario is further challenged by screen-shooting distortions. Therefore, a practical solution should preserve copyright information, enable source screen attribution, and resist watermark removal and forgery.


\section{Proposed Method}
\subsection{Problem Formulation and Overview}
Multi-screen collusion exploits the statistical redundancy among watermark residuals distributed across different screens. A captured watermarked image from screen $s$ can be modeled as
\begin{equation}
	I_w^s = I_o^s + R_s + N_s,
\end{equation}
where $I_o^s$, $R_s$, and $N_s$ denote the displayed image content, screen-specific watermark residual, and screen-shooting distortion, respectively.
Given captured copies $\{I_w^s\}_{s=1}^{K}$, an adversary estimates the residual by subtracting independently collected clean images $\{\tilde{I}_o^s\}_{s=1}^{K}$ and averaging:
\begin{equation}
	\hat{R}_{K}
	=
	\frac{1}{K}
	\sum_{s=1}^{K}
	(I_w^s-\tilde{I}_o^s).
\end{equation}

In content-diverse collusion, different screens display different images carrying the same copyright watermark, i.e., $I_o^1\neq I_o^2\neq\cdots\neq I_o^K$. The estimated residual can be decomposed as
\begin{equation}
	\hat{R}_{K}
	=
	\frac{1}{K}\sum_{s=1}^{K}R_s
	+
	\frac{1}{K}\sum_{s=1}^{K}(I_o^s-\tilde{I}_o^s)
	+
	\frac{1}{K}\sum_{s=1}^{K}N_s .
\end{equation}
Since content structures vary across different images, the averaged content-related residual is gradually reduced as the collusion scale $K$ increases:
\begin{equation}
	\frac{1}{K}
	\sum_{s=1}^{K}
	(I_o^s-\tilde{I}_o^s)
	\rightarrow 0,
	\quad K\rightarrow\infty .
\end{equation}
Therefore, the residual estimate gradually approaches the averaged watermark residual:
\begin{equation}
	\hat{R}_{K}
	\approx
	\frac{1}{K}\sum_{s=1}^{K}R_s+\epsilon ,
\end{equation}
where $\epsilon$ denotes mixed screen-shooting distortions.

In content-aligned collusion, all screens display the same image carrying the same watermark, i.e., $I_o^1=I_o^2=\cdots=I_o^K$. Since image structures remain consistent across colluded copies, they cannot be effectively suppressed through aggregation. The estimated residual can be represented as
\begin{equation}
	\hat{R}_{K}
	=
	\frac{1}{K}\sum_{s=1}^{K}R_s
	+
	R_{content}
	+
	\epsilon ,
\end{equation}
where $R_{content}$ denotes the preserved image structures and $\epsilon$ represents mixed screen-shooting distortions. 
Therefore, the adversary adjusts the residual strength factor $\alpha$ to control the influence of preserved image content on the collusion results.

The estimated residual is then exploited for watermark removal and forgery:
\begin{equation}
	I_{rem}=I_w-\alpha\hat{R}_{K},
	\qquad
	I_{for}=I_o+\alpha\hat{R}_{K}.
\end{equation}

As shown in Fig.~\ref{fig:framework}, CoMSMark consists of an image-agnostic screen-conditioned residual encoder $E_{nc}$, a screen-shooting noise layer $N$, and a dual-head decoder $D_{ec}$. Given watermark message $W$ and screen ID $S$, the encoder generates a screen-specific residual $I_r$ independent of the cover image $I_o$, enabling efficient large-scale distribution. The residual is added to $I_o$ to obtain the watermarked image $I_w$, which is then processed by $N$ to simulate screen-shooting distortions. Finally, the decoder $D_{ec}$ recovers the watermark message $W'$ and screen ID $S'$ through two task-specific heads.

\subsection{Screen-Conditioned Residual Embedding}

The goal of the screen-conditioned encoder $E_{nc}$ is to generate diverse watermark residuals for different screens while preserving the same copyright message. Given a watermark message $W$ and a screen ID $S$, the encoder learns a screen-conditioned mapping to produce a residual $I_r$. To achieve this, the screen ID is introduced as a conditional signal to modulate both channel characteristics and spatial distributions of watermark features.

Specifically, the watermark message $W$ is first projected by a linear layer and processed by ConvGNReLU ($3\times3$ convolution, group normalization, and ReLU activation) to obtain watermark features $F_w$. Meanwhile, the discrete screen ID $S$ is mapped into a continuous embedding vector through an embedding layer. This conditional vector is used for subsequent feature modulation and spatial fusion.

\subsubsection{Channel Modulation}
Channel modulation aims to control screen-specific feature characteristics of watermark residuals. Inspired by FiLM~\cite{perez2018film, wu2025versatile}, the screen embedding is processed by a channel MLP (ChaMLP) consisting of three linear layers and two LeakyReLU activations to generate modulation parameters $\{\gamma_1,\beta_1,\gamma_2,\beta_2\}$.

First, raw modulation is applied to the watermark feature $F_w$ to adjust channel responses:
\begin{equation}
	F_{mid}=F_w\odot\sigma(\gamma_1)+\beta_1 ,
\end{equation}
where $\odot$ denotes element-wise multiplication and $\sigma(\cdot)$ is the Sigmoid function.
Then, instance normalization is applied to refine the conditioned feature distribution:
\begin{equation}
	F_w'=\mathrm{IN}(F_{mid})\odot\sigma(\gamma_2)+\beta_2 .
\end{equation}

\subsubsection{Spatial Fusion}
Channel modulation controls feature characteristics but does not explicitly regulate spatial residual distribution. Therefore, we further introduce spatial fusion to generate screen-specific spatial patterns. The screen ID embedding is processed by a spatial conditioning network (SpaMLP) consisting of two linear layers and a LeakyReLU activation. Its output is reshaped into a spatial feature map $F_c$ and concatenated with $F_w'$. Three ConvGNReLU layers are then applied for nonlinear feature interaction, followed by an output convolution and Tanh activation to generate the final screen-conditioned residual $I_r$.
Since $I_r$ is generated independently of the cover image $I_o$, CoMSMark maintains an image-agnostic encoding paradigm~\cite{zhang2020udh}, avoiding repeated image-aware processing and enabling efficient large-scale distribution.


\subsection{Noise Layer}
The residual $I_r$ is linearly added to the cover image to obtain the watermarked image, which is then passed through a differentiable screen-shooting noise layer $N$. This layer simulates perspective transformation, illumination distortion, moiré patterns~\cite{fang2022pimog}, JPEG compression, Gaussian noise, Gaussian blur, and image scaling~\cite{zhu2018hidden,jia2021mbrs,tancik2020stegastamp}, producing the noised image $I_n$ for subsequent decoding.

\subsection{Dual-Head Watermark Extraction}
Given a noised image $I_n$, the decoder $D_{ec}$ aims to recover the watermark message $W'$ and predicts the corresponding screen ID $S'$. To improve feature sharing while preserving task-specific representations, we design a dual-head decoder consisting of a shared backbone, a watermark extraction head $W_{Head}$, and a screen identification head $S_{Head}$.

The shared backbone first extracts high-level features from $I_n$ using ConvGNReLU blocks and progressive downsampling. The $W_{Head}$ further refines the shared features with two residual blocks and projects them to the watermark dimension through fully connected layers. The $S_{Head}$ processes the shared features using a ConvGNReLU block and a residual block. Global average pooling and global max pooling are then applied in parallel to capture complementary global and salient screen-specific responses. The pooled features are concatenated and passed through an LNReLU block and fully connected layers to predict the screen ID.

By separating watermark recovery and screen identification into task-specific heads, the decoder jointly supports copyright authentication and leakage source attribution under screen-shooting distortions in multi-screen scenarios.

\subsection{Training Strategy}

Training CoMSMark involves several interdependent and potentially conflicting objectives, including visual fidelity, watermark robustness, screen identification, and collusion resistance. Improving robustness may affect visual quality, while introducing screen-specific diversity should not interfere with watermark decoding. To stabilize optimization, we progressively introduce the loss terms over four training stages. The model first learns watermark decoding, then incorporates visual constraints, and finally introduces screen ID supervision and collusion suppression.

In the first stage, covering the initial third of training, only watermark decoding loss is optimized:
\begin{equation}
	L_{dec\_w}=\left\|W^{'}-W\right\|^2 .
\end{equation}

In the second stage, spanning one-third to one-half of training, we introduce a visual loss that combines pixel-level MSE and LPIPS~\cite{zhang2018unreasonable}:
\begin{equation}
	L_{mse}=
	\left\|
	\gamma(I_w)-\gamma(I_o)
	\right\|^2 ,
\end{equation}
\begin{equation}
	L_{enc}
	=
	\lambda_1L_{mse}
	+\lambda_2L_{lpips}(I_w,I_o),
\end{equation}
where $\gamma(\cdot)$ denotes a differentiable RGB-to-YUV transformation, and $\lambda_1=1$, $\lambda_2=0.1$.

In the third stage, screen ID supervision is introduced using cross-entropy loss:
\begin{equation}
	L_{dec\_s}=L_{ce}(S^{'},S).
\end{equation}

\begin{figure*}[!htb]
	\centering
	\includegraphics[width=0.8\linewidth]{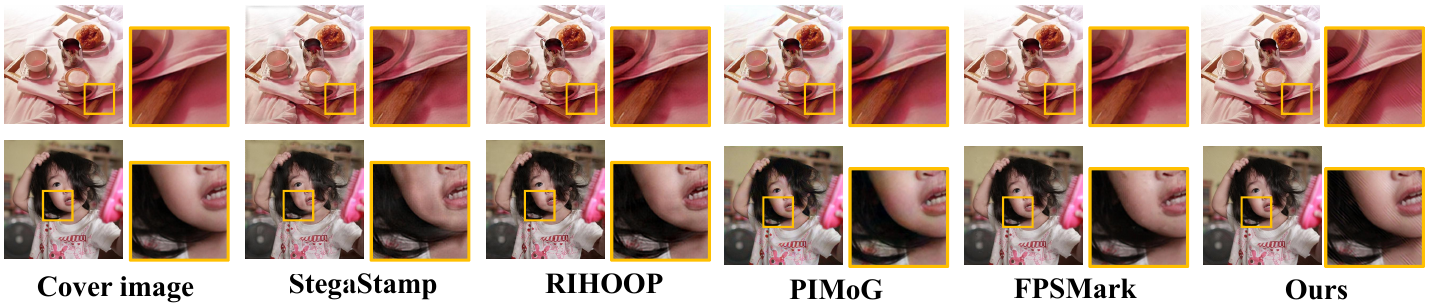}
	\caption{Visual comparison of watermarked images generated by different methods. Enlarged views of the marked regions are provided for detailed inspection of embedding artifacts.}
	\label{fig3}
\end{figure*}

In the final stage, we introduce the collusion suppression loss. For each mini-batch, the same watermark $W$ is paired with different screen IDs $\{S_i\}_{i=1}^{B}$ to generate residuals $\{R_i\}_{i=1}^{B}$. The mean residual is computed as
\begin{equation}
	\bar{R}=\frac{1}{B}\sum_{i=1}^{B}R_i ,
\end{equation}
and suppressed by
\begin{equation}
	L_{msl}=\|\bar{R}\|_2^2 .
\end{equation}
To simulate collusion forgery, we add $\bar{R}$ to clean images:
\begin{equation}
	\tilde{I}_o=I_o+\bar{R}.
\end{equation}
The forged samples are decoded to obtain watermark probabilities $p_{col}\in[0,1]^{B\times L}$. Since binary prediction entropy reaches its maximum at 0.5, we encourage uncertain predictions by
\begin{equation}
	L_{adl}
	=
	\frac{1}{BL}
	\|p_{col}-0.5\|_1 .
\end{equation}

The overall collusion suppression loss is
\begin{equation}
	L_{coll}
	=
	\lambda_3L_{msl}
	+\lambda_4L_{adl},
\end{equation}
where both $\lambda_3$ and $\lambda_4$ are set to 1.

The final objective is
\begin{equation}
	\mathcal{L}
	=
	\lambda_wL_{dec\_w}
	+\lambda_eL_{enc}
	+\lambda_sL_{dec\_s}
	+\lambda_{coll}L_{coll}.
\end{equation}
After activation, the weights are set to $\lambda_w=10$, $\lambda_e=5$, $\lambda_s=2$, and $\lambda_{coll}=0.5$. This progressive optimization strategy reduces interference among competing objectives and stabilizes training.

\begin{table}[!t]
	\centering
	\setlength{\tabcolsep}{0.7mm}
	\begin{tabular}{cccccc}
		\toprule
		Method & StegaStamp & RIHOOP & PIMoG & FPSMark & Ours\\
		\midrule
		PSNR &27.23 &31.41 &32.99 &\textbf{35.93} & 35.85\\
		SSIM &0.810 &0.903 &0.941 &\textbf{0.952} & 0.949\\
		\bottomrule
	\end{tabular}
	\caption{The PSNR and SSIM values of each method.\label{tab:table1}}
	\label{table:table1}
\end{table}

\section{Experimentation}
\subsection{Implementation Details}
We randomly select 10,000 images from COCO~\cite{lin2014microsoft} for training and 1,000 images for testing. All images are center-cropped and resized to $400\times400$ pixels. The watermark is a random binary sequence of length $L=100$, and the screen ID is sampled from 128 identities, indexed from 0 to 127. Within each mini-batch, the same watermark is paired with different screen IDs to generate screen-specific residuals for collusion-aware training. CoMSMark is implemented in PyTorch and trained on an NVIDIA RTX 4090 GPU using the AdamW optimizer~\cite{loshchilov2017decoupled} with a learning rate of $5\times10^{-5}$ and a batch size of 16. The progressive training schedule lasts 300 epochs.

For screen-shooting evaluation, we randomly select 100 test images and compare CoMSMark with four representative methods: StegaStamp~\cite{tancik2020stegastamp}, PIMoG~\cite{fang2022pimog}, RIHOOP~\cite{jia2020rihoop}, and FPSMark~\cite{chen2025flexible}. Watermark robustness is measured by bit accuracy (ACC), while visual quality is evaluated using peak signal-to-noise ratio (PSNR) and structural similarity index measure (SSIM). Screen attribution is measured by Screen Identification Accuracy (SID-ACC), defined as the Top-1 accuracy of the predicted screen ID. 
For collusive removal, PSNR is computed between the captured watermarked image and its attacked counterpart, whereas for collusive forgery, it is computed between the clean image and the forged image.
Unless otherwise specified, an iPhone 12 is used for capture and a Lenovo XiaoXin Pro 14 is used for display. The phone is fixed on a tripod and remotely triggered via Bluetooth, and all methods are evaluated under the same capture conditions.

\subsection{Visual Quality}
We evaluate the visual quality of watermarked images using PSNR and SSIM, with quantitative results reported in Table~\ref{table:table1} and qualitative comparisons shown in Fig.~\ref{fig3}. CoMSMark achieves a PSNR of 35.85 dB and an SSIM of 0.949, demonstrating visual quality comparable to FPSMark and competitive performance among existing methods. These results demonstrate that CoMSMark preserves image fidelity while generating screen ID conditioned watermark residuals for source attribution.

As shown in Fig.~\ref{fig3}, StegaStamp introduces visible artifacts in smooth regions, while PIMoG exhibits slight color shifts. In contrast, CoMSMark produces watermarked images that remain visually close to the cover images, with fewer perceptible artifacts and a more natural appearance.

\begin{figure}[!t]
	\centering
	\includegraphics[width=\linewidth]{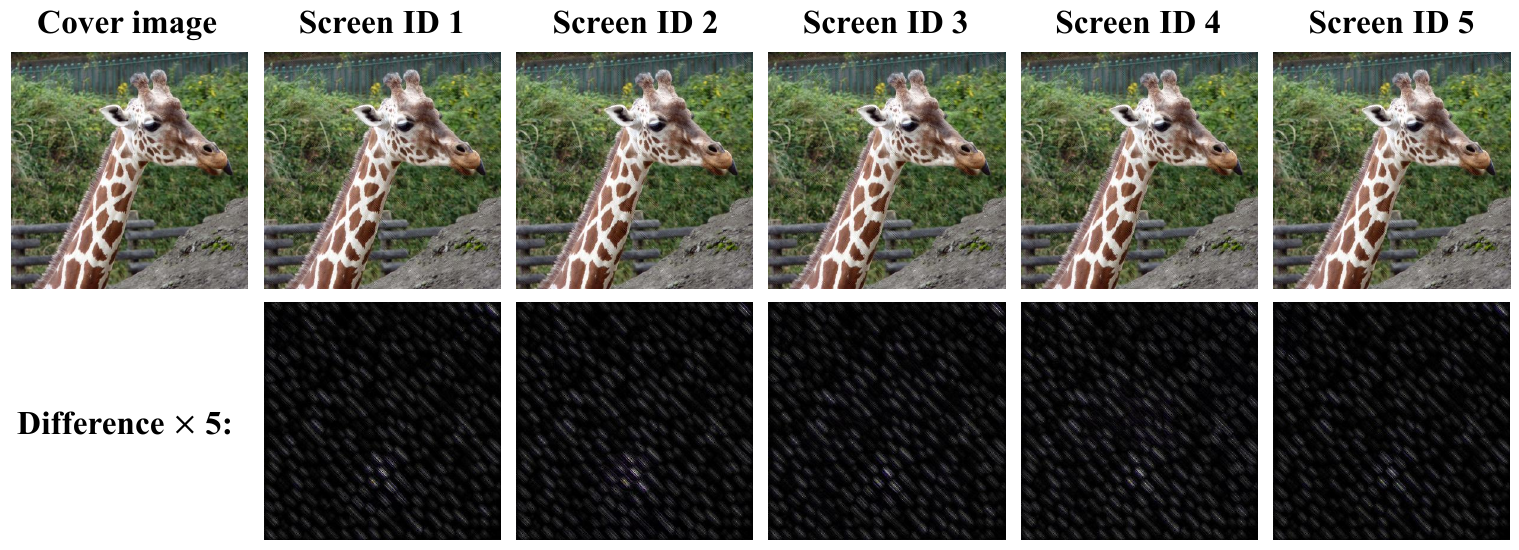}
	\caption{Screen-specific watermark residuals generated from the same copyright watermark.}
	\label{fig:4}
\end{figure}

\subsection{Multi-Screen Watermark Decoding Performance}
We evaluate whether CoMSMark can simultaneously support reliable copyright watermark recovery and screen ID attribution from captured images. Two complementary settings are considered: a fixed watermark paired with different screen IDs, and different watermarks paired with the same screen ID. As reported in Table~\ref{table:table2}, both ACC and SID-ACC exceed 99\% in the two settings. These results show that screen ID conditioning does not compromise watermark recovery, while variations in the watermark message have little effect on screen attribution. The dual-head decoder therefore reliably separates copyright recovery from leakage source identification.

Fig.~\ref{fig:4} further visualizes residuals generated from the same watermark under different screen IDs. The residuals retain similar global structures associated with the shared copyright information while exhibiting noticeable screen-specific variations in their spatial distributions. This indicates that screen ID conditioning preserves watermark consistency while introducing sufficient cross-screen diversity. Such a balance supports reliable screen identification and helps suppress shared residual components exploitable by collusion attacks.

\begin{table}[!t]
	\centering
	\setlength{\tabcolsep}{4mm}
		\begin{tabular}{ccc}
			\toprule
			Setting & ACC & SID-ACC \\
			\midrule
			Fixed $W$, Varying $S$ & 99.74\% & 100\%\\
			Fixed $S$, Varying $W$ & 99.96\% & 100\%\\
			\bottomrule
		\end{tabular}
		\caption{Watermark and screen ID decoding performance under different configurations.\label{tab:table2}}
		\label{table:table2}
	\end{table}

\subsection{Collusion Resistance Evaluation}

We evaluate CoMSMark against collusive watermark removal and forgery under two scenarios. In content-diverse collusion, different images carry the same watermark but different screen IDs, and the collusion scale is varied. In content-aligned collusion, the same image is assigned different screen IDs. Since residual averaging may preserve image textures, we fix the collusion scale at 50 and vary the collusion strength. The estimated residual is subtracted from watermarked images for removal and added to clean images for forgery. ACC measures attack effectiveness, while PSNR evaluates the visual quality of attacked images. Quantitative PSNR results are reported in Table~\ref{table:table3}, with visual examples of content-diverse collusion shown in Fig.~\ref{fig:7}. Additional visual results are provided in Section.A of the supplementary material.

\begin{table*}[t]
	\centering
	\setlength{\tabcolsep}{2.7mm}
	\begin{tabular}{llccccc|ccccc}
		\toprule
		\multirow{2}{*}{Attack}
		& \multirow{2}{*}{Method}
		& \multicolumn{5}{c|}{Content-Diverse Collusion (Scale)}
		& \multicolumn{5}{c}{Content-Aligned Collusion (Strength)} \\
		\cmidrule(lr){3-7}
		\cmidrule(lr){8-12}
		&
		& 20 & 40 & 60 & 80 & 100
		& 0.2 & 0.4 & 0.6 & 0.8 & 1.0 \\
		\midrule
		
		\multirow{5}{*}{\makecell[l]{Collusive\\Watermark\\Removal}}
		& StegaStamp
		& 17.94 & 21.26 & 21.38 & 22.42 & 23.35
		& 26.35 & 20.33 & 16.81 & 14.31 & 12.37 \\
		& RIHOOP
		& 17.92 & 21.36 & 21.51 & 22.59 & 23.62
		& 26.32 & 20.30 & 16.78 & 14.28 & 12.34 \\
		& PIMoG
		& 17.99 & 21.36 & 21.48 & 22.51 & 23.49
		& 26.32 & 20.30 & 16.78 & 14.28 & 12.34 \\
		& FPSMark
		& 18.81 & 22.28 & 22.30 & 23.25 & 24.31
		& 26.47 & 20.45 & 16.93 & 14.43 & 12.49 \\
		& Ours
		& 19.16 & 21.37 & 21.84 & 23.19 & 23.39
		& 26.27 & 20.25 & 16.73 & 14.23 & 12.29 \\
		\midrule
		
		\multirow{5}{*}{\makecell[l]{Collusive\\Watermark\\Forgery}}
		& StegaStamp
		& 18.36 & 21.84 & 21.92 & 23.03 & 24.03
		& 26.56 & 20.69 & 17.32 & 15.00 & 13.25 \\
		& RIHOOP
		& 18.34 & 21.95 & 22.07 & 23.22 & 24.32
		& 26.53 & 20.66 & 17.30 & 14.98 & 13.23 \\
		& PIMoG
		& 18.40 & 21.95 & 22.04 & 23.16 & 24.23
		& 26.53 & 20.66 & 17.30 & 14.97 & 13.22 \\
		& FPSMark
		& 19.15 & 22.82 & 22.83 & 23.89 & 25.04
		& 26.67 & 20.79 & 17.43 & 15.10 & 13.34 \\
		& Ours
		& 19.55 & 21.82 & 22.37 & 23.84 & 24.04
		& 26.48 & 20.61 & 17.24 & 14.92 & 13.17 \\
		\bottomrule
	\end{tabular}
	\caption{PSNR (dB) under collusive watermark removal and forgery.}
	\label{table:table3}
\end{table*}
\begin{figure}[!t]
	\centering
	\includegraphics[width=\linewidth]{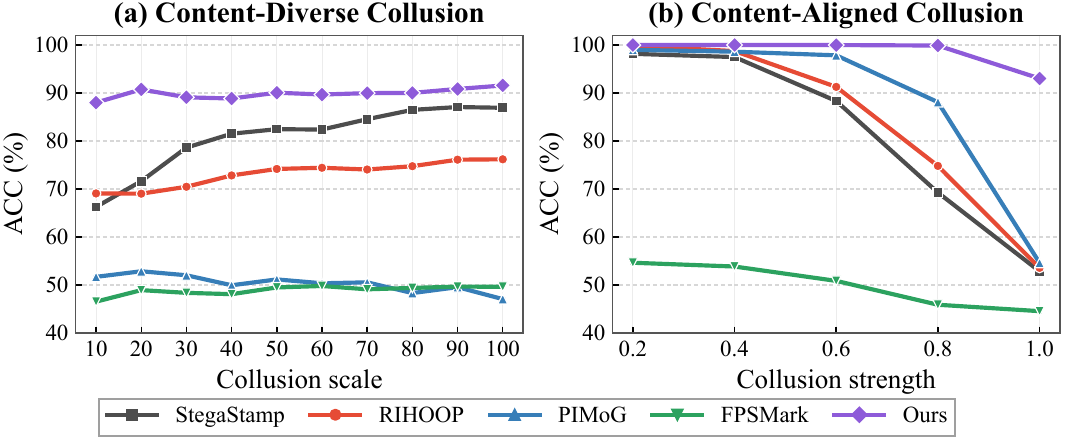}
	\caption{Watermark decoding accuracy under collusive watermark removal.}
	\label{fig:5}
\end{figure}

\subsubsection{Collusive Watermark Removal}
For collusive watermark removal, an independently sampled clean image is subtracted from each captured watermarked image to obtain a residual estimate. The clean and watermarked images are not paired. The resulting residual estimates are averaged to approximate the shared watermark residual, which is then subtracted from the captured watermarked images. In the content-diverse collusion scenario, the collusion scale ranges from 10 to 100 screen IDs. In the content-aligned collusion scenario, the collusion scale is fixed at 50, and the mean residual is scaled by a coefficient ranging from 0.2 to 1.0 before subtraction. The ACC and PSNR results are reported in Fig.~\ref{fig:5} and Table~\ref{table:table3}, respectively.

In the content-diverse collusion scenario, increasing the collusion scale progressively attenuates the image-specific textures retained in the estimated residual. CoMSMark consistently maintains an ACC of approximately 90\%, while the competing methods exhibit substantially lower decoding accuracy. In particular, PIMoG and FPSMark remain close to random guessing under most settings. Meanwhile, the PSNR of the attacked images generally increases with the collusion scale, indicating that larger collusion sets produce visually cleaner removal results. These results show that CoMSMark preserves strong watermark recoverability even under large-scale collusive removal.

In the content-aligned collusion scenario, increasing the removal strength reduces PSNR from approximately 26 dB to 12 dB, indicating severe visual degradation. The ACCs of StegaStamp, RIHOOP, and PIMoG decrease sharply and approach random guessing at the highest strength. In contrast, CoMSMark retains an ACC of 93.03\% at a removal strength of 1.0, demonstrating strong resistance to collusive watermark removal.



\begin{figure}[!t]
	\centering
	\includegraphics[width=\linewidth]{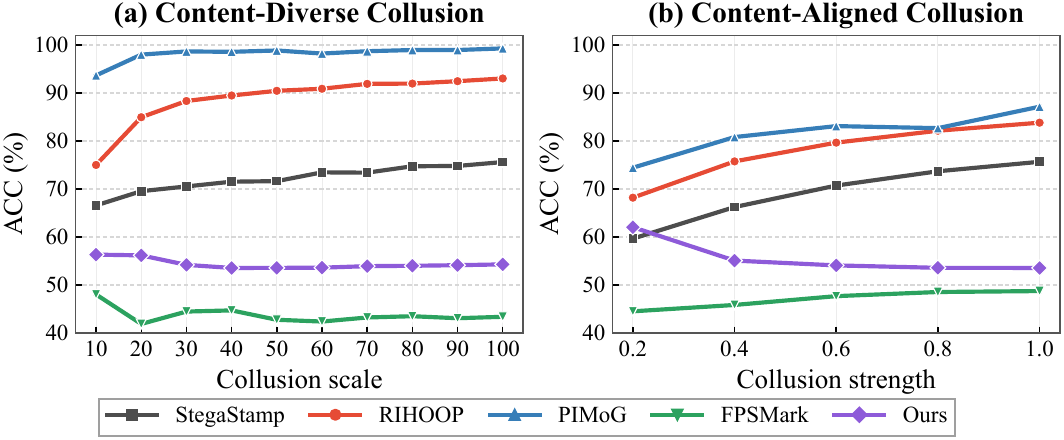}
	\caption{Watermark decoding accuracy under collusive watermark forgery.}
	\label{fig:6}
\end{figure}

\subsubsection{Collusive Watermark Forgery}
For watermark forgery, the estimated mean residual is added to clean images to construct forged samples. A higher watermark ACC on the forged images indicates a more successful forgery attack, whereas an ACC close to 50\% indicates stronger resistance to collusive forgery. The ACC and PSNR results are reported in Fig.~\ref{fig:6} and Table~\ref{table:table3}, respectively.

In the content-diverse collusion scenario, the forgery ACCs of PIMoG, RIHOOP, and StegaStamp increase with the collusion scale, reaching 99.30\%, 93.03\%, and 75.60\%, respectively, at a scale of 100, which indicates weak resistance to collusive forgery. By contrast, CoMSMark remains stable between 53\% and 56\% across all scales, staying close to random guessing and achieving performance comparable to FPSMark. Meanwhile, the PSNR of the forged images increases with the collusion scale, showing that larger collusion sets produce visually cleaner forged samples. Thus, CoMSMark remains resistant to collusive watermark forgery even when the attacker uses a large collusion set.

A similar trend is observed in the content-aligned collusion scenario. As the attack strength increases, the forgery ACCs of PIMoG, RIHOOP, and StegaStamp rise substantially, whereas CoMSMark remains close to random guessing, decreasing from 62.02\% at a strength of 0.2 to 53.53\% at a strength of 1.0. At the same time, PSNR drops from approximately 26 dB to 13 dB as the attack strength increases, revealing a clear trade-off between forgery effectiveness and image usability. These results confirm that CoMSMark effectively resists collusive watermark forgery across different attack strengths.



\begin{figure}[!t]
	\centering
	\includegraphics[width=\linewidth]{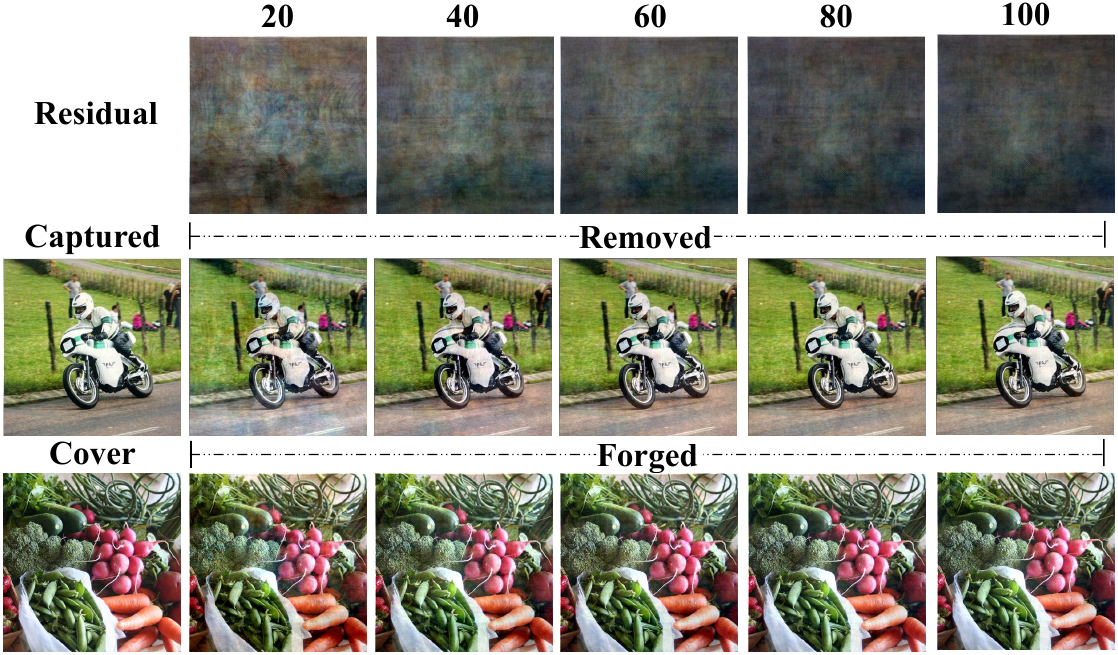}
	\caption{Visual examples of colluded images under different collusion scales in the content-diverse scenario.}
	\label{fig:7}
\end{figure}

\subsection{Robustness Under Different Capture Conditions}
We evaluate robustness under variations in capture distance, angle, and device combination. As reported in Table~\ref{table:table4}, all methods achieve high watermark decoding accuracy across the tested distances and angles, with the angular results averaged over the left and right capture directions. CoMSMark and FPSMark show the most stable performance, both maintaining ACC above 99\% under all settings. Across different device combinations, CoMSMark achieves ACC above 98\% and SID-ACC above 96\%, demonstrating strong robustness to practical hardware variations. Detailed results are provided in Section.B of the supplementary material.

\begin{table}[t]
	\centering
	\setlength{\tabcolsep}{2.5pt}

	\begin{tabular}{lccc|ccc}
		\toprule
		\multirow{2}{*}{Method}
		& \multicolumn{3}{c|}{Distance (cm)}
		& \multicolumn{3}{c}{Angle ($^\circ$)} \\
		\cmidrule(lr){2-4}
		\cmidrule(lr){5-7}
		& 20 & 30 & 40
		& 20 & 30 & 40 \\
		\midrule
		StegaStamp
		& 99.65 & 99.86 & 99.85
		& 99.84 & 99.90 & 99.77 \\
		
		RIHOOP
		& 99.77 & 99.83 & 99.81
		& 99.83 & 99.75 & 99.76 \\
		
		PIMoG
		& 99.43 & 98.40 & 98.53
		& \textbf{100.00} & 99.89 & 99.84 \\
		
		FPSMark
		& 99.54 & \textbf{100.00} & \textbf{99.93}
		& \textbf{100.00} & 99.90 & \textbf{99.90} \\
		
		Ours
		& \textbf{99.92} & 99.93 & 99.90
		& 99.95 & \textbf{99.92} & 99.88 \\
		\bottomrule
	\end{tabular}
	\caption{Watermark decoding accuracy under different capture distances and angles. Angle results are averaged over the left and right directions.}
	\label{table:table4}
\end{table}

\section{Conclusion}

This paper presents a multi-screen collusion attack against screen-shooting watermarking and proposes CoMSMark to defend against it. By conditioning residual generation on screen ID, CoMSMark produces screen-specific residuals for source attribution without increasing watermark capacity. A collusion suppression loss further reduces shared residual components and improves resistance to collusive removal and forgery. Its image-agnostic encoder also enables efficient large-scale distribution. Extensive experiments show that CoMSMark outperforms state-of-the-art methods under both attacks while maintaining competitive visual quality and robustness under practical screen-shooting conditions.

\bibliography{aaai2027}


\end{document}